\documentclass[aps,showpacs,twocolumn,superscriptaddress,nofootinbib,floatfix,section 10pt]{revtex4-2}
\usepackage{amsfonts,amssymb,stmaryrd,latexsym,amsmath,braket}
\usepackage{graphicx,subfigure}
\usepackage{comment}
\usepackage{times}
\usepackage{slashed}
\usepackage{bm}
\usepackage{braket}
\usepackage{appendix}
\usepackage{enumitem}
\usepackage{multirow}
\usepackage{array}

\usepackage[colorlinks=true,backref=true,linktocpage=true,citecolor=blue,urlcolor=blue,linkcolor=blue,pdfpagemode=UseOutlines]{hyperref}
\usepackage[dvipsnames]{xcolor}

\begin{document}
\title{Gauge invariance and color charge fluctuations}
\author{Thomas D. Cohen}
\email{cohen@umd.edu}
\affiliation{Department of Physics and Maryland Center for Fundamental Physics, University of Maryland, College Park, MD 20742 USA}

\begin{abstract}
The nature of confinement is connected with color charge.  Unfortunately, the color charge densities in QCD, the Noether charge densities associated with the global color invariance, are not invariant under local color rotations.  This implies that the expectation values of the net color charge in any region of any physical state in QCD--states that satisfy the color Gauss law--are automatically zero for all components of color.   In this paper it is shown that the expectation value of the square of the net color charge in a region---a measure of the color charge fluctuations---is necessarily nonzero when evaluated in physical states and the result, while depending on the scheme and scale by which the theory is regulated, is gauge invariant.  This holds despite the formal lack of gauge invariance of the operator. Moreover, there is a particular combination of the color charge fluctuations for the vacuum and for a system describable by a non-trivial density matrix that is independent that has a well-defined continuum limit. 
\end{abstract}

\date{\today}
\maketitle
\section{Introduction}
Understanding the nature of confinement has been a central theoretical problem in QCD from its earliest days\cite{Wilson:1974sk}.  Indeed, precisely what is meant by ``confinement''\cite{Greensite:2011zz} is not totally clear, beyond the very basic fact that QCD does not have isolated states with net color: there are no isolated quarks or gluons, so it is natural to think of the issue as one of color confinement.  A related issue of phenomenological significance is that at sufficiently high temperatures, the thermodynamics of QCD acts like that of a plasma of quarks and gluons;  this suggests that the quarks and gluons enter a deconfined quark-gluon plasma  phase\cite{Cabibbo:1975ig}.  While  Cabbibo and Parisi's original idea that there the quark-gluon plasma is a distinct deconfined phase is not quite right, the notion of a quark-gluon plasma has become central in nuclear physics  and  has led to a massive experimental effort in heavy ion physics to determine its  properties (for a review see \cite{Pasechnik:2016wkt}) .    There have been numerous approaches to understanding the notion of confinement in QCD (or Yang-Mills theory) since the earliest days of the theory\footnote{These include approaches aimed at understanding issues of confinement via a linear rising potential and the area law of the Wilson loop\cite{Wilson:1974sk,Kogut:1974ag}, the role of center symmetry and the role of center symmetry breaking  and the Polyakov   loop\cite{Polyakov:1975rs,Polyakov:1978vu}, approaches based on 't Hooft loops, the vertex free energy, or center vertices \cite{tHooft:1977nqb, tHooft:1979rtg}, dual superconductors \cite{Mandelstam:1974vf,Mandelstam:1974pi} or the Kugo-Ojima approach\cite{Kugo:1979gm}.  These and other approaches were extensively studied in the decades since their formulation; the state of the art circa 2010 was reviewed in ref.~\cite{Greensite:2011zz}}.  Theoretical attempts to understand confinement are ongoing and new approaches are still being developed---such as a recent scheme describing deconfinement via  percolation of electric center fluxes\cite{Ghanbarpour:2022oxt}.  

It is probably fair to say that there is no real consensus on how confinement works or even what it means to understand confinement.  However,  the most basic notion of confinement is some variant of the idea that particles carrying a color charge cannot be isolated.  It is quite easy to show  that gauge invariance requires the expectation value of every component of the net color charge in any spatial region to be zero for any physical state in the theory.  However, confinement in the most basic sense clearly means something more than that: imagine a hypothetical isolated quark or gluon in a state where the direction in color-space has quantum fluctuations  such that average color in any direction is zero but where the square of the of the net color charge---a measure of the fluctuations---in a spatial region surrounding the quark to be nonzero.  Such a state should not be possible in the physical spectrum of a confined theory.   Thus, understanding color charge fluctuations may be a key to gaining insights on confinement\footnote{Understanding color charge fluctuations is of importance for other aspects of strong interaction physics.  See for example ref.~\cite{PhysRevD.70.105012}}.  

This paper focuses on QCD---the theory of relevance to the standard model---but applies to other confining gauge theories.  The discussion  uses a Hamiltonian formulation based on a lattice regularization of the theory\cite{Kogut:1974ag}.  Ultraviolet regularization is central to the analysis, and the lattice is the most natural way to do this. Similarly, the color Gauss law plays an essential role; it is natural to view this as a restriction on the physical states in a Hamiltonian formulation. The analysis is implicitly in the Scr\"odinger picture where the time dependence is in the states rather than the operators. A necessary first step  is gauge fixing.  Although the results do not depend on the choice of gauge, for the sake of concreteness, it is helpful to think of the analysis as being in the temporal gauge with $A_0^a=0$.   This choice yields a simple connection between the vector potential and the color electric field  and a simple Hamiltonian\cite{Kogut:1974ag},  making it a natural gauge choice for possible treatments of lattice QCD in future quantum computers\cite{Banuls:2019bmf,Lamm:2019uyc,Cohen:2021imf,Bauer:2022hpo,Bauer:2023qgm}.  

Gauge fixing does not exhaust the gauge freedom; there is residual gauge freedom associated with purely spatial gauge transformation.  Since spatial gauge transformations do not alter the physical state, physical states must be invariant under these  transformations.   The Euler-Lagrange equations of the theory include equations without time dependence, which in the continuum are given by: 
 \begin{equation}
 \begin{split}
G^a(\vec{x})&= 0 \; \;  {\rm where}\\ G^a(\vec{x}) &\equiv\vec \nabla \cdot \vec{E}^a(\vec{x})-  g \, \rho^a_{\rm color} (\vec{x}) \; ,
\end{split}
 \end{equation}
where $g$ is the coupling constant (and is implicitly defined at the ultraviolet scale at which the theory is defined). This is the color Gauss law.  The Gauss-law operators act as generators of spatial gauge transformations, which constrains the class of physical states: physical states are invariant under residual gauge transformations---if $|\psi \rangle_{\rm phys}$ is physical, then $G^a(\vec{x})|\psi \rangle_{\rm phys}=0$ for all $a$ and all $\vec{x}$. The physical Hilbert space is smaller than the full Hilbert space of theory.

Understanding color confinement is complicated since the color charge density associate with the color $a^{th}$ component $SU(N_c)$ global symmetry, given in the continuum by,
\begin{equation}
\rho^a_{\rm color}(\vec{x})=f^{a b c} A_i^b(\vec{x}) E_i^c(\vec{x}) \,  + \, \sum_f \overline q_f(\vec x) \gamma_0 \lambda^a q_f(\vec{x})
\end{equation}
where $f$ is the  flavor and  $E_i^a$,  the color electric field ({\it i.e|} the $0^{\rm th}$ component of the field strength tensor), 
is not gauge invariant and non-gauge invariant objects are typically thought of as unphysical.  
 Moreover, invariance of physical states under residual gauge transformations implies trivially that for all physical states $ {}_{\rm phys}\langle \phi| \rho^a_{\rm color}(\vec{x}) |\psi \rangle_{\rm phys} =0 $.

However, as will be shown in this paper, despite the non-gauge-invariant nature of the color charge and the triviality of the color charge matrix elements, there are gauge invariant----and hence physical---measures of color charge fluctuations in spatial regions.  Even though the net global color charge of any physical state is zero, this does not rule out,  nonzero color charges in different spatial regions with correlations yielding a vanishing net color but nonzero color fluctuations.  An analogy with isopin might be helpful.  A theory with exact isospin can have a state $|\psi \rangle $ that is isosinglet ($I^2=0$) but which has two spatial regions, each with isospin content.  If  the isospin  of the two regions are $\vec I_1$ and $\vec I_2$ with $\vec I =\vec I_1 +\vec I_2$, then the Wigner-Eckert theorem implies that $\langle \psi|\vec I_1 |\psi \rangle = \langle \psi|\vec I_2 |\psi \rangle =0$.  However, this does not imply that there is no isospin in each region, but  rather that the isospin in each region can fluctuate in a correlated way with 
 $\langle \psi|\vec I_1^2| \psi \rangle=\langle \psi|\vec I_2 ^2| \psi \rangle = - \langle \psi|\vec I_1 
 \cdot  \vec I_2 + \vec I_2 
 \cdot  \vec I_1 |\psi \rangle$.  That is, while the expectation value of every component of the isospin in each regions is zero, the fluctuations of isospin need not be.
 
Global color symmetry is analogous to isospin and this paper focuses on the  color charge fluctuations in a spatial region, $R$. The operator, given by 
 \begin{equation}
     {\cal Q}^2_R =\sum_a \left (\int_R d^3 x  \rho^a_{\rm color}(\vec{x}) \right )^2 \; ,
 \end{equation}
is a color singlet (and, thus not forced to vanish by the Wigner-Eckert theorem); its expectation value yields the color fluctuations. However, unlike with isospin there is also local gauge symmetry and ${\cal Q}^2_R$ 
is not gauge invariant\footnote{An easy way to see that it is not gauge invariant is to consider a configuration in which the color charge density is non-zero only in two spatially disjoint regions denoted $1$ and $2$; in both regions the color charge density points in a single color direction $b$:  $\rho^a_{\rm color}=\rho(\vec{x}) \delta_{a b}$ with $\rho(\vec{x})=0$ unless $\vec{x}$ is in region 1 or 2. Consider a region $R$ that fully contains both regions 1 and 2; then   ${\cal Q}^2_R$ has contributions from region 1, region 2 and a non-zero cross term.  Next consider a gauge transformation which is unity over region 1 and is a constant color rotation over region 2 to a direction orthogonal to $b$.  The contribution to ${\cal Q}^2_R$ from region 1 and 2  will remain unchanged under this transformation, but the cross term vanishes.}.  It is often assumed that unless an operator is gauge invariant, it is unphysical, which might suggest that ${\cal Q}^2_R$ is not physically relevant.  However, the assumption that non-gauge-invariant operators are necessarily unphysical is not quite right. 

Let $\cal O$ be an operator without explicit time dependence or dependence on the time component of the gluon field operator, $A_0$. Suppose further that it is not invariant under pure spatial gauge transformations: $U^\dagger O U \ne O$ where $U$ is a unitary operator corresponding to a spatial gauge transformation. Using ${\cal P}$, the projection operator onto the physical subspace,  ${\cal O}$ can be decomposed into two parts, one, ${\cal O}_{\rm GI} \equiv {\cal P O P}$, is purely gauge invariant while the other ${\cal O}_{\rm GV}  = {\cal O} - {\cal O}_{\rm GI}$ is purely gauge variant;  By construction, matrix elements of ${\cal O}_{\rm GV}$ between the physical states are zero---as is expected from an operator lacking physical content. On the other hand, if ${\cal O}_{\rm GI} \equiv {\cal P O P} \ne 0$, then $ {}_{\rm phys}\langle \phi| {\cal  O }  |\psi \rangle_{\rm phys} =  {}_{\rm phys}\langle \phi| {\cal P O P}  |\psi \rangle_{\rm phys} \; $ is generically non-zero and is independent of gauge transformations; {\it i.e.} it corresponds to something physical.  Thus, if ${\cal P}{\cal Q}^2_R {\cal P} \ne 0$, it matrix elements fixed spatial regions are physical.  The projectors are onto states that satisfy the color Gauss law, which implies that
\begin{widetext}
\begin{equation}
     {}_{\rm phys}\langle \phi| {\cal Q}^2_{ R } |\psi \rangle_{\rm phys} = \left (
    \frac 1 {g^2}\right) \,
\sum_a \int_{\partial R} d^2 x \int_{\partial R} d^2 y  \,  {}_{\rm phys}\langle \phi|\left ( \hat{n}(\vec x) \cdot \vec E^a (\vec x) \right )  \, \left ( \hat{n}(\vec y) \cdot \vec E^a (\vec y) \right)|\psi \rangle_{\rm phys}
\end{equation}
\end{widetext}
where $\partial R$ is the surface bounding $R$ and $\hat{n}$ is the outward normal. Of course, $\sum_a  \left ( \hat{n}(\vec x) \cdot \vec E^a (\vec x) \right )  \, \left ( \hat{n}(\vec y) \cdot \vec E^a (\vec y) \right)$ is, in general, not gauge invariant.  It is gauge-invariant if, and only if,  $\vec{x}= \vec{y}$ and  the double integral only has contributions when $\vec{x}= \vec{y}$:
\begin{subequations}
\begin{equation}
\begin{split}
&{\cal P} \int_{\partial R} d^2 y ( \hat{n}(\vec x) \cdot \vec E^a (\vec x)  \left ( \hat{n}(\vec y) \cdot \vec E^a (\vec y) \right) {\cal P} \\& \, = \, 
a_l^2 {\cal P}  \left ( \hat{n}(\vec x) \cdot \vec E^a (\vec x)  \right ) ^2  {\cal P} \; \;  {\rm so \, that}\\
& {}_{\rm phys}\langle \psi| {\cal Q}^2_{ R } |\psi \rangle_{\rm phys} =  \\
& {}_{\rm phys}\langle \psi|  \int_{\partial R} d^2 x \left (\frac{a_l}{g(a_l)} \hat{n}(\vec x) \cdot \vec E^a (\vec x)  \right ) ^2  |\psi \rangle_{\rm phys}
\end{split}
\end{equation}
where $a_l$ is a constant with dimensions of length and $g(a_l)$ is the coupling constant at that length scale.  The factor of $a_l^2$ follows from dimensional analysis, which requires that whatever regulates the implicit $\delta$ function  has dimensions of length squared.  It is often interesting to consider expectation values in ensembles of states and to focus on $\langle {\cal Q}^2_{ R } \rangle_{\rho_{\rm phys}}$, the expectation value of color fluctuation in a system described by $\rho_{\rm phys}$ a density matrix restricted to physical states.
\begin{equation}
\begin{split}
&\langle {\cal Q}^2_{ R } \rangle_{\rho_{\rm phys}}=\\
& \left( \frac{a_l}{g(a_l)}\right)^2 \,  {\rm Tr} \left [\rho_{\rm phys}  \int_{\partial R}\!\!\!d^2 x \sum_a \left ( \hat{n}(\vec x) \cdot \vec E^a (\vec x)  \right ) ^2 \right ] \ ,
\end{split}\end{equation}
where the trace is over states in the physical Hilbert space.
\end{subequations}

Physically, the factor of $a_l^2$  arises since the continuum quantum theory is ill-defined;  some ultraviolet regulator is needed to define the theory. The coupling constant is fixed at the scale at which the theory is defined.  The regulator restricts the distance scale over which the electric field is allowed to vary, which, in turn,  restricts the allowable class of spatial gauge transformations.   If the theory is regularized via a cubic lattice, $a_l^2$ is the area of one face of a cell and the integrals over space are replaced by sums:  
\begin{equation}
\begin{split}
&\langle {\cal Q}^2_{ R } \rangle_{\rho_{\rm phys}} =\\
&\left ( \frac{a_l}{g(a_l)} \right )^2  {\rm Tr} \left [\rho_{\rm phys}\sum_{k \in \partial R} 
 \sum_{f} \sum_a \left (\vec E_k^a \cdot \hat n_k^f \right )^2\right ]
 \label{Eq:qsq}
\end{split} \end{equation}
where $k$ represents lattice sites  on the boundary of the region and $f$ a face on that cell which forms the boundary (so that  $\hat n_f^k$ represents the unit vector orthogonal to that face).    The analysis goes through, {\it mutatis mutandis}, for other choices of regulator. The operator $\sum_{k \in \partial R} 
 \sum_{f} \sum_a \left (\vec E_k^a \cdot \hat n_k^f \right )^2 $ on the  right-hand side of Eq.~ (\ref{Eq:qsq}) is gauge invariant; thus, the value of $\langle {\cal Q}^2_{ R } \rangle_{\rho_{\rm phys}} $ is unchanged by gauge transformations on the operator---it is gauge invariant---even though the operator on which it is based, $ {\cal Q}^2_{ R }$, is not.  This is a principal result of this paper.  

One might be concerned that the dependence on $a_l^2$ means that$ \langle {\cal Q}^2_{ R } \rangle_{\rho_{\rm phys}}$ is zero in the continuum limit.  This is not the case. Consider the vacuum state.  As $a_l^2 \rightarrow 0$,  $\vec E_a)^2$ can be computed perturbatively:    $\sum_a (\hat n \cdot \vec E_a)^2 \rightarrow\frac{ \rm const}{a_l^4} $, where the constant can depend on the details of the regulator.   This implies that in vacuum, ${\cal Q}^2_{ R }$, is
\begin{equation}
\langle {\cal Q}^2_{ R } \rangle_{\rm vac} = 
\frac{ 4 \pi  \, {\rm const}^2 A_R}{ a_l^2 \, \alpha_s(a_l) }
\end{equation}\\
where $A_R$ is the area of the region. Far from vanishing in the continuum limit, this quantity diverges.  

The result seems counterintuitive: in the absence of long-range correlations, the mean-square fluctuation of a conserved quantity in a translationally invariant system scales with the system's volume.  Instead, $\langle {\cal Q}^2_{ R } \rangle_{\rm vac}$  scales with the area.  This is surprising and implies large correlations in the color charge.

\begin{figure}[t]
    \includegraphics[width=.39 \textwidth]{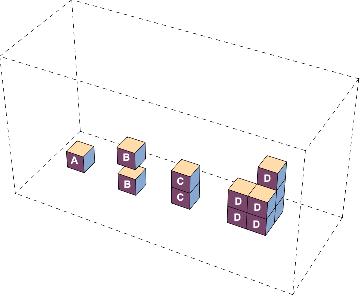}
    \caption{Each block is cubic and with the same volume.  Region A is a single block, Region B has two disconnected blocks, Region C has two connected blocks with one common face, region D has 7 connected blocks with a total of 8 faces shared by two distinct blocks .}
    \label{fig:VacFig}
\centering 
\end{figure}

Since  $\langle {\cal Q}^2_{ R } \rangle_{\rm vac} 
 \rightarrow \infty$  as $a_l \rightarrow 0$ it is not well defined in the continuum limit. 
This issue does not arise if one considers the relative values of the charge fluctuation in different spatial regions of the vacuum rather than their absolute levels. For two regions $R_1$ and $R_2$
\begin{equation}
\frac{\langle {\cal Q}^2_{ R_1 } \rangle_{\rm vac} }{\langle {\cal Q}^2_{ R_2 } \rangle_{\rm vac} } = \frac{A_{R_1}}{A_{R_2}}
\end{equation}
regardless of the value of  $a_l$.  The peculiar nature of the color charge fluctuations in the vacuum is illustrated by the four regions in Fig.~\ref{fig:VacFig}.  The ratio of the volumes of the four regions and the ratio of the square of the vacuum charge fluctuations are given by 
\begin{equation}
\begin{split}
    &V_A:V_B:V_c:V_D = 1:2:2:7 \; \; {\rm while}\\
   & \langle Q^2_{R_A}\rangle_{\rm vac}:\langle Q^2_{R_B}\rangle_{\rm vac}:\langle Q^2_{R_C}\rangle_{\rm vac}:    \langle Q^2_{R_D}\rangle_{\rm vac}\\& = 1:2:\frac53:\frac{13}3 \; .
    \end{split}
\end{equation}
 As noted earlier,  the fluctuations do not scale with volume:  region D is 7 times the volume of region A, but the color charge fluctuations are only $\frac{ 13}{3}$ times larger.  Regions B and C illustrate a related peculiarity: the fluctuations in region B are larger than in C by a factor of $\frac65$ due to the gap in the disconnected regions of $B$.  That the ratio remains $\frac65$ even if the gap size is altered but if the regions merge, the ratio becomes unity. 
    
Next, consider color fluctuations when the system is not the vacuum but rather is described by a non-trivial density matrix. Unfortunately, this is of little utility directly: as $a_l \rightarrow 0$, all states will be dominated by the ultraviolet, which is the same as in the vacuum contribution and diverges.  This suggests a focus on the change in the color fluctuations relative to what is there in the vacuum:
\begin{equation}
\begin{split}
&\left (\Delta {\cal Q}_{ R } \right )^2 \equiv \sum_a\left ( \int_R d^3 x  \rho^a_{\rm color}(\vec{x}) -\sqrt{\frac{\langle{\cal Q}_ R  ^2\rangle_{\rm vac}}{N_c^2-1}} \right )^2  \; \;  \\
& {\rm so \; that}\;  \; \langle  \, \left (\Delta {\cal Q}_{ R } \right )^2 \, \rangle_{\rho_{\rm phys}}
= {\langle \cal Q}_ R^2 \rangle_{\rho_{\rm phys}}  -  \langle{\cal Q}_R  ^2\rangle_{\rm vac}\; .
\end{split}
\end{equation}
Unfortunately,  the expectation value of $\Delta {\cal Q}^2_{ R }$ is proportional to $a_l^2$ 
and vanishes in the continuum limit.

Fortunately, there is a quantity, 
that depends  solely on $\langle  \,  {\cal Q}_{ R }^2 \, \rangle_{\rm vac}$ and $ \langle  \, \left (\Delta {\cal Q}_{ R } \right )^2 \, \rangle_{\rho_{\rm phys}}$  that is well-defined and nontrivial in the continuum limit:
\begin{equation}
\begin{split}
&F_{R ; \rho _{\rm phys}} \equiv \\
 &\lim_{a_l \rightarrow 0} \,  \log
\left ( \langle  \,  {\cal Q}_{ R }^2 \, \rangle_{\rm vac} \right ) \sqrt{\langle  \,  {\cal Q}_{ R }^2 \, \rangle_{\rm vac} \langle  \, \left (\Delta {\cal Q}_{ R } \right )^2 \, \rangle_{\rho_{\rm phys}} } = \\&
   \lim_{a_l \rightarrow 0} \,  c \Lambda^2 {\rm Tr} \left [\rho_{\rm phys}  \int_{\partial R} d^2 x \,  \sum_a\left (\hat{n}(\vec x) \cdot \vec E^a (\vec x)  \right ) ^2   \right. \\&
  \; \; \; \; \; \; \; \; \; \; \; \; - \; \left . \rho_{\rm phys}\,  A_R \, \frac{\left (E_{\rm vac}\right )^2}{3}  \right ]  
\end{split}
\end{equation}
$\Lambda^2$ is the square of the mass scale of the theory ({\it eg.} $\Lambda_{\rm QCD}^2$) and $c$ is a dimensionless constant that depends on the $\beta$ function and  details of how the theory is regularized.
Up to logarithmic corrections, $F_{R ; \rho _{\rm phys}}$ is simply the geometric mean of $\langle  \,  {\cal Q}_{ R }^2 \, \rangle_{\rm vac}$ and $\langle  \, \left (\Delta {\cal Q}_{ R } \right )^2 \, \rangle_{\rho_{\rm phys}} $.

Consider a purely thermal density matrix characterized by temperature, $T$, with no additional constraints.  In that case, the ratio of $F_{\rho _{T}}^R$ in distinct regions, like the vacuum case, is directly proportional to $A_R$, the area of the region.  Thus, for the regions in Fig.~\ref{fig:VacFig}, 
\begin{equation}
    F_{R_A; T}:F_{R_B;T}:F_{R_C; T}:F_{R_D; T} = 1:2:\frac53:\frac{13}3 \; .
\end{equation}
For a single region, the ratio of $F_{R_A; T}$ at two temperatures is 
\begin{equation}
\frac{ F_{R; T_1}}{ F_{R; T_2}}= \frac{\langle E^2 \rangle_{T_1} - \langle E^2 \rangle_{\rm vac}}{\langle E^2 \rangle_{T_2} - \langle E^2 \rangle_{\rm vac}}  \; \underset{T_1,T_2 \rightarrow \infty}{\longrightarrow} \; \frac{T_1^2}{T_2^2} \; ;
\end{equation}
the arrow represents the high temperature limit with $\frac{T_1}{T_2}$ fixed.

As a final example,  consider the ground state of Yang-Mills theory with two well-separated fundamental color charges ({\it ie.~}arbitrarily heavy quarks) added to the system in a gauge-invariant way via a Wilson line.  The behavior of such a system has long been known\cite{Wilson:1974sk,Kogut:1974ag, Creutz:1980zw,Bali:2000gf,Greensite:2011zz}: close to the sources, the color electric fields are Coulomb-like---$ \left(\vec E^2 - \langle \vec E^2 \rangle_{\rm vac} \right )$ falls off as $1/r^4$---up logarithmic corrections; well away from the sources, a flux tube (responsible for a linear rising potential) forms.   A small spherical region centered on one of the charges has $F_{R ; \rho _{\rm phys}}$ falling  off as $1/r^2$ (up to logarithmic corrections).  This might appear to suggest that  $F_{R ; \rho _{\rm phys}}$ represents a color charge localized at the source with some form of color screening.  However, the interpretation of  $F_{R ; \rho _{\rm phys}}$ is not viable. Considers cylindrical region coaxial with the flux tube.  Recall that well away from well-separated charges, the flux tube has a more or less constant width  with a color electric field strength that is approximately independent of the distance to the charges.  For a cylindrical region coaxial with the flux tube and wide enough to fully contain it, one sees that $F_{R ; \rho _{\rm phys}}$ is non-zero and has a value that is approximately independent of the length of the cylinder and where it is placed between the sources.   This implies shifts in the color fluctuations from the vacuum in regions well away from the sources is nonzero and independent of the size of the region;  this looks nothing like color charges localized at the source with some form of color screening.   

In conclusion,  fluctuations of the color charge evaluated in physical states are gauge invariant,  despite the lack of gauge-invariance of the associated operator.  Moreover, a combination of the vacuum expectation value of the color fluctuations and their shift from the vacuum   is independent of the ultraviolet regularization scale.   The peculiar behavior of  $ \langle {\cal Q}^2_{ R } \rangle_{\rho_{\rm phys}}$, might prompt one  to query whether it ``really'' should be thought of as the expectation value of color fluctuations.   That question of interpretation might be dismissed as being entirely of a philosophical rather than scientific nature.  However, an important question remains open: Can studies of this quantity ultimately provide useful insights into the physics of confinement?

\begin{acknowledgments}

This work was supported in part by the U.S. Department of Energy, Office of Nuclear Physics under Award Number(s) DE-SC0021143, and DE-FG02-93ER40762.

\end{acknowledgments}
  \bibliography{main.bib}
\end{document}